\documentclass[12pt]{article}

\usepackage[english]{babel}

\usepackage[letterpaper,top=2.5cm,bottom=2.5cm,left=2.5cm,right=2.5cm,marginparwidth=1.75cm]{geometry}
\usepackage{float}
\usepackage{authblk}

\usepackage{comment}
\usepackage{amsmath}
\usepackage{amsthm}
\usepackage{amsfonts}
\usepackage{graphicx}
\usepackage{csquotes}
\usepackage{booktabs}
\usepackage{multirow}
\usepackage[colorlinks=true, allcolors=blue]{hyperref}
\usepackage[round]{natbib}
\newtheorem{prop}{Proposition}
\title{Coarsened data in small area estimation: a Bayesian two-part model for mapping smoking behaviour}
\author[1]{Aldo Gardini}
\author[1]{Lorenzo Mori}

\affil[1]{Department of Statistical Sciences ``Paolo Fortunati'', University of Bologna, Italy}

\date{}

\begin{document}
\maketitle

\begin{abstract}\noindent
Estimating health indicators for restricted sub-populations is a recurring challenge in epidemiology and public health. When survey data are used, Small Area Estimation (SAE) methods can improve precision by borrowing strength across domains. In many applications, however, outcomes are self-reported and affected by coarsening mechanisms, such as rounding and digit preference, that reduce data resolution and may bias inference. This paper addresses both issues by developing a Bayesian unit-level SAE framework for semi-continuous, coarsened responses. Motivated by the 2019 Italian European Health Interview Survey, we estimate smoking indicators for domains defined by the cross-classification of Italian regions and age groups, capturing both smoking prevalence and intensity. The model adopts a two-part structure: a logistic component for smoking prevalence and a flexible mixture of Lognormal distributions for average cigarette consumption, coupled with an explicit model for coarsening and top-coding. Simulation studies show that ignoring coarsening can yield biased and unstable domain estimates with poor interval coverage, whereas the proposed model improves accuracy and achieves near-nominal coverage. The empirical application provides a detailed picture of smoking patterns across region–age domains, helping to characterize the dynamics of the phenomenon and inform targeted public health policies.
\end{abstract}

\section{Introduction}

A substantial proportion of empirical research in public health is grounded in sample surveys, which routinely collect self-reported information on health behaviors, risk factors, and lifestyle patterns. Although such surveys are an essential source of evidence for population health monitoring, the resulting data are inherently subject to sampling variability and multiple forms of measurement error \citep[see, among others,][]{schenker2010improving, ramo2011reliability}. Self-reported behavioral outcomes are particularly prone to recall bias, rounding, and systematic digit preference, all of which can distort descriptive analysis and both design- and model-based inference. Furthermore, survey sample sizes are often insufficient to produce reliable estimates for small geographic domains or specific socio-demographic groups, thereby motivating the development and use of statistical techniques capable of \enquote{borrowing strength} across areas.

This study focuses on tobacco consumption, a leading modifiable risk factor for which reliable and geographically disaggregated measures are essential for monitoring population health. To this end, we use data from the European Health Interview Survey (EHIS), coordinated by Eurostat, focusing on the Italian 2019 component. The key indicators considered in the analysis are computed from the variable \enquote{mean number of cigarettes smoked daily} and are potentially affected by data coarsening in self-reports, which reduces the effective resolution of the observed data. In particular, we focus on estimating the proportion of daily smokers, the average daily number of cigarettes smoked among daily smokers, and the proportion of heavy smokers among daily smokers. Furthermore, a key objective of this work is to produce estimates at the regional level and further disaggregated by age group, which exacerbates the problem of limited sample sizes in several domains.

Producing reliable regional and age-specific estimates is important for informing targeted public health policies and resource allocation \citep{kong2020use, flor2021effects}. From a statistical perspective, it also highlights why standard direct estimators, that is expansion estimators based solely on the sample, may be unreliable for the smoking indicators of interest. In particular, direct estimators treat self-reported daily cigarette consumption as error-free, so systematic coarsening can propagate into biased domain-level quantities such as proportions and conditional means. At the same time, the limited sample sizes inflate sampling variability, making direct estimates unstable even in the absence of measurement error. Taken together, these features suggest that direct estimation alone may be inadequate for the targets of interest.

Coarsening, heaping, and digit preference have long been recognized as sources of bias in epidemiology, demography, and survey statistics. These phenomena can be interpreted within a unified conceptual framework of data coarsening, whereby rounding represents the most elementary mechanism through which continuous values are observed only at a reduced resolution, while heaping corresponds to structured forms of coarsening characterized by the concentration of reported values at preferred digits or multiples, possibly with multiple levels of coarseness coexisting within the same dataset \citep{heitjan1990inference}. Seminal methodological contributions by \citet{heitjan1991ignorability} and \citet{heitjan1993ignorability,heitjan1994ignorability} extend the concept of ignorability from missing data to more general forms of data coarsening and formalize the conditions under which the stochastic coarsening mechanism can be validly ignored. When the coarsening mechanism depends on the value of the response, however, ignorability no longer holds and the coarsening process must be modeled explicitly, a point that is particularly relevant for self-reported outcomes subject to heaping.

Building on this framework, subsequent work explicitly models heaping and other non-random coarsening mechanisms, showing that failure to account for such processes can substantially distort parameter estimates and lead to misleading inference \citep{zhang2007impact, wang2008modeling, zinn2016statistical}. More recent contributions examine the impact of heaping in regression modeling \citep{bar2012accounting} and propose adjustment strategies, possibly incorporating additional information from validation data \citep{ahmad2024method}. Empirical studies document substantial heaping in behavioral measurements \citep{crawford2015sex} and demographic counts \citep{camarda2017modelling}, while survey-oriented work emphasizes implications for official statistics indices such as the poverty rate \citep{drechsler2016beat}.

Despite this extensive literature, most methodological work on coarsened data has focused on individual-level outcomes, whereas the main targets in official statistics and public health monitoring are domain-level indicators. This shift from individual to domain targets is central when interest lies in small geographic areas or specific population subgroups. Small Area Estimation (SAE) provides a natural framework to produce stable domain estimates by borrowing strength across areas through auxiliary information and random effects \citep{rao_small_2015}. While SAE methods are well developed, the integration of explicit coarsening mechanisms within SAE models has received limited attention.

We address this gap with a Bayesian unit-level SAE model that embeds an explicit coarsening mechanism, treating reported consumption values as heaped realizations of an underlying latent continuous smoking intensity and estimating the coarsening process jointly with the small area model parameters. This joint formulation propagates coarsening-related uncertainty into the target domain indicators while still exploiting auxiliary information for precision gains. The data structure motivates two additional components, a two-part specification to accommodate the non-negligible mass at zero \citep{dreassi2014small, chandra2016small}, and a mixture of two Lognormal components for the positive part to capture heterogeneity in daily cigarette consumption. This specification also raises a technical issue that is relevant for inference on domain indicators. Because Bayesian models based on a Lognormal likelihood may yield posterior moments that are not finite under commonly used priors \citep{gardini2022poverty}, we also derive sufficient conditions, stated in terms of prior assumptions on the variance components, that guarantee the existence of posterior moments for the target indicators. Existing SAE applications to tobacco outcomes \citep{hermes2012small, eberth2018estimating, wang2022spatial, santiago2023small} either focus on prevalence type measures only, or, when addressing intensity, do not account for systematic coarsening. Our model-based simulation study in Section~\ref{sec:4} further shows that ignoring coarsening can substantially deteriorate estimator performance, leading to biased point estimates and degraded uncertainty quantification.

The paper is organized as follows. Section~\ref{sec:2} introduces the data and the motivation for the study. Section~\ref{sec:3} presents the proposed two-part SAE model for coarsened data. Section~\ref{sec:4} describes a model-based simulation study comparing alternative specifications. Section~\ref{sec:5} presents the empirical application on smoking behaviour in Italy. Section~\ref{sec:6} concludes the paper.

\section{The data and the motivation}\label{sec:2}

The EHIS is a harmonized survey conducted across EU member States to produce comparable indicators on population health, health care use, and health determinants \citep{hintzpeter2019european}. In Italy, EHIS 2019 was implemented by the Italian National Institute of Statistics (Istat). Data were collected from approximately 22,800 households in 837 municipalities through face-to-face interviews and self-administered questionnaires, covering individuals aged 15 and over. The sampling design is multi-stage, with municipalities as primary sampling units and households as secondary units, and is stratified by municipality size.

Our analysis focuses on smoking behavior in Italy, with domains defined by the 21 NUTS 2 regions and four age groups, for a total of 84 region--age domains. We consider three key indicators: (i) the proportion of daily smoking, (ii) the average number of cigarettes smoked per day among daily smokers, and (iii) the proportion of heavy smokers among daily smokers, defined as those who smoke at least 20 cigarettes per day. These indicators are widely used for public health monitoring and are regularly reported by institutions such as Eurostat and the American Lung Association \citep{Eurostat2024_smokers, AmericanLungAssociation2024_smoking}.

\begin{figure}[]
    \centering
    \includegraphics[width=1\linewidth]{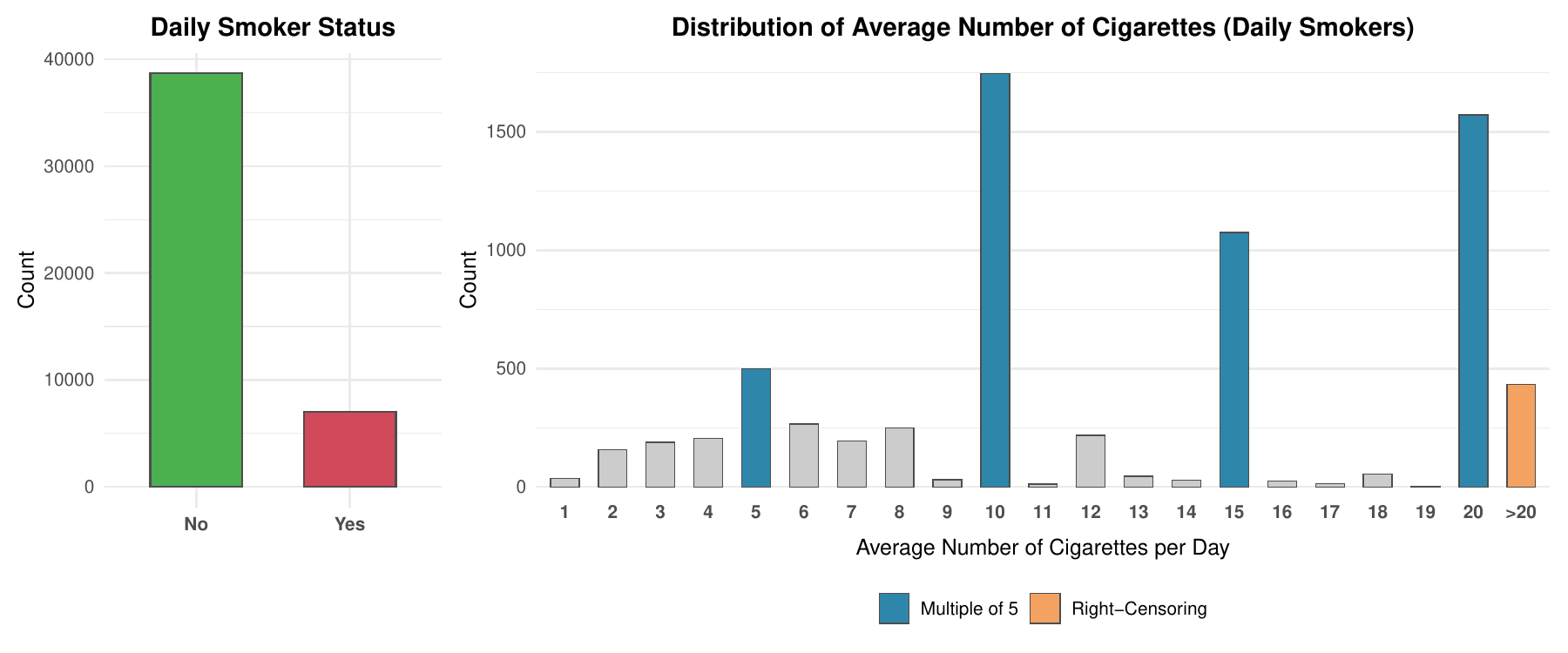}
    \caption{Bar-plots of Daily smoker status (left panel) and distribution of average number of cigarettes smoked by daily smokers (right panel).}
    \label{fig:fig1}
\end{figure}

In EHIS, smoking behavior is collected through a skip pattern: respondents are first asked about current smoking status and, among those who report smoking daily, the survey records the average number of cigarettes smoked per day. Although the underlying quantity can be viewed as continuous, it is recorded on a discrete scale (an integer count) and subject to top-coding, with all values above 20 grouped into a single category. For the purposes of this study, we define the daily number of cigarettes as zero for individuals who are not daily smokers, while positive values are observed only among daily smokers. Figure~\ref{fig:fig1} illustrates the resulting data features: the left panel shows the predominance of zero values, whereas the right panel displays the distribution of recorded consumption among daily smokers. The pronounced spikes at 5, 10, 15, and 20 cigarettes, together with the upper category pooling values above 20, indicate strong rounding and digit preference, as well as right-censoring induced by top-coding.

These measurement features have direct implications for estimation. Standard design-based estimators implicitly treat the recorded number of cigarettes as error-free, but, in this setting, positive values are coarsened by rounding/heaping and by top-coding. As a result, indicators that depend on the intensity component, such as the mean number of cigarettes among daily smokers and the proportion of heavy smoking, can be distorted when computed directly from the reported counts, because intermediate values tend to pile up at preferred numbers and all consumption above 20 is collapsed into a single category. By contrast, the proportion of daily smoking relies on the binary classification of respondents as daily smokers or not, and is therefore not affected by these reporting issues.

Our goal is to estimate the three smoking indicators for each of the 84 region-age domains. In many domains, however, the sample size is limited, with a minimum of 106 observations and an average of 544.3 per domain, implying that direct domain estimators are unstable and potentially highly variable. This naturally motivates the use of SAE methods, which improve precision by borrowing strength across related domains through statistical modeling. A common SAE option is the area-level approach, where the model is built on domain-specific direct estimates. In our case, this strategy is not feasible because, for indicators involving cigarette consumption intensity, direct domain estimates are either unavailable or not reliable: reported counts exhibit strong heaping at preferred values and are top-coded above 20 cigarettes per day, so the within-domain information needed to identify mean consumption and the proportion of heavy smoking is often too limited or distorted. We therefore adopt a unit-level SAE approach, which models individual outcomes and leverages auxiliary information observed for the full population. In our application, population registers provide covariates such as sex, education, and age group, which are included in the unit-level model to produce coherent estimates for all 84 domains. The next section formalizes these ideas by introducing the notation and the quantities used throughout the paper.

\subsection{Notation and target quantities}\label{sec:target}

Consider a finite population $U$ of size $N$, partitioned into $D$ domains
$U_1,\ldots,U_D$ with sizes $N_1,\ldots,N_D$, such that $N=\sum_{d=1}^D N_d$.
A sample $s$ of size $n$ is selected from $U$ according to a possibly complex sampling design.
Let $s_d$ denote the domain-specific sample with size $n_d$, where $0\le n_d\le N_d$
and $\sum_{d=1}^D n_d=n$.

Let $Y$ denote the individual-level average number of cigarettes smoked per day.
To capture the semicontinuous nature of smoking intensity, we decompose
$
Y = W\cdot Z,
$
where $W$ is the indicator of daily smoking,
\[
W=
\begin{cases}
1 & \text{if the individual is a daily smoker,}\\
0 & \text{otherwise,}
\end{cases}
\]
and $Z\in\mathbb{R}^+$ denotes the average number of cigarettes smoked per day among daily smokers, noting that the value of $Z$ is not relevant when $W=0$.
Unit-level values are denoted by $y_{id}=w_{id}z_{id}$ for $i=1,\ldots,N_d$ and $d=1,\ldots,D$.
Auxiliary information is available for each unit and is collected in the $p$-dimensional vector
$\mathbf{x}_{id}=(x_{id1},\ldots,x_{idp})^\top$, $\forall i,d$.

In this context, the following domain indicators are typically considered:
\begin{itemize}
    \item the proportion of daily smoking in domain $d$:
    \begin{equation}\label{eq:w_d}
    \overline{w}_d = \frac{1}{N_d}\sum_{i=1}^{N_d} w_{id};
    \end{equation}

    \item the mean daily number of cigarettes among daily smokers:
    \begin{equation}\label{eq:z_d}
    \bar{z}_d = \frac{\sum_{i=1}^{N_d} w_{id} z_{id}}{\sum_{i=1}^{N_d} w_{id}},
    \qquad \text{assuming } \sum_{i=1}^{N_d} w_{id}>0;
    \end{equation}

    \item the proportion of heavy smoking among daily smokers, defined as $Z\geq20$:
    \begin{equation}\label{eq:HS_d}
    \overline{HS}_d = \frac{\sum_{i=1}^{N_d} w_{id}\,\boldsymbol{1}_{[20,\infty)}\!\left(z_{id}\right)}
    {\sum_{i=1}^{N_d} w_{id}},
    \qquad \text{assuming } \sum_{i=1}^{N_d} w_{id}>0,
    \end{equation}
    where $\boldsymbol{1}_A(e)$ is an indicator function assuming value $1$ if $e\in A$ and 0 otherwise.
\end{itemize}


\section{A two-part small area model for heaped data}\label{sec:3}

This section presents the proposed unit-level Bayesian small area model, formulated in a two-part framework to reflect the semi-continuous nature of daily cigarette consumption \citep{dreassi2014small, chandra2016small}. Section \ref{sec:part1} introduces the logistic mixed model for the probability of being a daily smoker. Section \ref{sec:part2} then develops the smoking intensity component, treating consumption as a latent continuous quantity observed with reporting error arising from coarsening and top-coding in the questionnaire. We point out that the two parts of the modeling strategy are completely separated, given that it is possible to factorize the likelihood of the models. Section \ref{sec:p3} discusses technical details on model fitting and checking, while Section \ref{sec:p4} defines the hierarchical Bayes (HB) predictors for the target functionals. Finally, Section \ref{sec:prior} introduces the prior structure, completing the hierarchical specification and paying particular attention to ensuring the existence of posterior moments for relevant quantities.

\subsection{Part one: daily smoker proportion}\label{sec:part1}
The model assumed for the dichotomous component $w_{id}$ is a typical logistic regression model with random effects. More in detail, a Bernoulli distribution with parameter $\nu_{id}=\mathbb{P}[w_{id}=1|\mathbf{x}_{id}]$ is assumed
$$
w_{id}|\nu_{id}\stackrel{ind}{\sim}\text{Ber}\left(\nu_{id}\right),\quad i=1,\dots,N_d,\ d=1,\dots,D.
$$
A linear predictor is specified for the logit transformation of $\nu_{id}$
$$
\text{logit}\left(\nu_{id}\right)=\log\left(\frac{\nu_{id}}{1-\nu_{id}}\right)=\beta_0^\nu+\mathbf{x}_{id}^\top\boldsymbol{\beta}^\nu+u_d^\nu,\quad \forall i,d,
$$
where $\beta_0^\nu\in\mathbb{R}$ is the intercept, $\boldsymbol{\beta}^\nu\in\mathbb{R}^p$ is the vector of regression coefficients and $u_d^\nu$ is a domain-specific random effect for which we assume the prior
$
u_d^\nu|\tau_\nu\stackrel{ind}{\sim}\mathcal{N}\left(0,\tau_\nu^2\right),\ \forall d.
$
To ease notation in the following sections, let the vector $\boldsymbol{\delta}\in\boldsymbol{\Delta}$ contain all the parameters involved in this modeling part. Being in the Bayesian setting, the model needs to be completed by setting appropriate prior distributions for the parameters. This task will be discussed in Section \ref{sec:prior}. 

\subsection{Part two: smoking intensity}\label{sec:part2}
The second part of the model is devoted to modeling smoking intensity among daily smokers, i.e. it targets the variable $Z\in\mathbb{R}^+$ conditional on $W=1$. Unfortunately, the self-reported data do not allow us to specify a direct model for this variable. Indeed, as already discussed in Section~\ref{sec:2}, several issues must be addressed. First, the survey requires an integer response for the average number of cigarettes smoked per day, which introduces an intrinsic rounding mechanism. Second, we observe a common phenomenon affecting self-reported data, known in the literature as coarsening, heaping, or digit preference. Lastly, the available data are top-coded, as all values larger than 20 are recorded as 21, leading to right censoring. Hence, our goal is to make inference on the distribution of $Z$, denoted by $f_Z(z|\boldsymbol{\theta})$ and governed by a vector of unknown parameters $\boldsymbol{\theta}\in\boldsymbol{\Theta}$. However, we only dispose of observations of the manifest variable $Z^\star$. To complete the setting, we introduce an additional latent variable $G$, which we assume to follow a conditional distribution $f_{G|Z}(g|z,\boldsymbol{\gamma})$ governed by $\boldsymbol{\gamma}\in\boldsymbol{\Gamma}$ and depending on $Z$. The role of $G$ is to describe the heaping behavior that directly affects the manifest variable $Z^\star$. We remark that the model discussed in this section is tailored to the applied setting considered in the paper; however, the underlying ideas can be transferred to other situations.

The latent variable $G$ plays a crucial role in the model formulation and we assume that it can take the following distinct values:
$$
G\in\mathcal{G}=
\begin{cases}
    1 \text{ if the respondent rounds to the nearest integer,}\\
    5 \text{ if the respondent rounds to the nearest multiple of 5,}\\
    10 \text{ if the respondent rounds to the nearest multiple of 10.}
\end{cases}
$$
Concerning the distribution of the latent variable, we assume a categorical distribution on the elements of $\mathcal{G}$, defining the probabilities as
$$
\mathbb{P}[G=g|z,\boldsymbol{\gamma}]=\lambda_g(z,\boldsymbol{\gamma}),\quad g\in\mathcal{G}.
$$
These probabilities depend on the latent variable $Z$ and we assume a proportional odds model as in \citet{wang2008modeling}:
\begin{equation}\label{eq:probsG}
\begin{cases}
    \lambda_1(z,\boldsymbol{\gamma}) = \text{expit}(\gamma_{01}+\gamma_1\log(z)),\\
    \lambda_5(z,\boldsymbol{\gamma}) = \text{expit}(\gamma_{02}+\gamma_1\log(z))-\text{expit}(\gamma_{01}+\gamma_1\log(z)), \\
    \lambda_{10}(z,\boldsymbol{\gamma}) = 1-\text{expit}(\gamma_{02}+\gamma_1\log(z)),\\
\end{cases}
\end{equation}
with the constraint $\gamma_{01}<\gamma_{02}$.
In addition, the coarsening regime determined by $G$, together with the values assumed by the manifest variable $Z^\star$, imply different ranges of values for the latent counterpart $Z$, defined as:
$$
\begin{cases}
    I_1(z^\star)=\begin{cases}
     (0,1.5) & \text{if } z^\star = 1\\
     [z^\star-0.5,\, z^\star+0.5) & \text{otherwise;}
\end{cases}\\
I_5(z^\star) = [z^\star-2.5,\, z^\star+2.5),\quad \text{if } z^\star\in\{5,10,15,20\};\\
    I_{10}(z^\star) = [z^\star-5.5,\, z^\star+4.5),\quad \text{if } z^\star\in\{10,20\}.
\end{cases}
$$

The observed coarsened answer $Z^\star$ is determined by the pair of latent variables as
$
Z^\star=T_{Z^\star}(Z,G)
$;
even if certain values of $Z^\star$ can be produced by multiple combinations of $Z$ and $G$. For example, $Z^\star=3$ is obtained from $T(Z\in I_1(3),\, G=1)$, whereas $Z^\star=10$ is obtained by three possible combinations: $T(Z\in I_1(10),\, G=1)$, $T(Z\in I_5(10),\, G=5)$, or $T(Z\in I_{10}(10),\, G=10)$. To formalize this latter feature, \citet{heitjan1994ignorability} introduced a further data element, again determined by the latent variables through a transformation $H=T_{H}(Z,G)$, that contains the information about what is known about the coarsening process from the data. This quantity is a set containing all the values that $G$ could take. Referring to the previous examples, $H=\{1\}$ if $Z^\star = 3$ and $H = \{1,5,10\}$ if $Z^\star = 10$.

As the data variables $Z^\star$ and $H$ are functions of the latent variables $Z$ and $G$, their conditional distribution given the latter is degenerate and it is 
\begin{equation}\label{eq:con_joint}
    f_{Z^\star,H|Z,G}(z^\star_{id},h_{id}|z_{id},g_{id}) = \boldsymbol{1}_{I_{g_{id}}(z_{id}^\star)}(z_{id})\boldsymbol{1}_{h_{id}}(g_{id}).
\end{equation}
From a practical viewpoint, the likelihood marginalized with respect to the latent variables is required and it is defined as
\begin{equation}\label{eq:lik_init}
    f_{Z^\star,H}(z^\star_{id},h_{id}|\boldsymbol{\theta,\boldsymbol{\gamma}})=\int_{\mathbb{R}^+}\sum_{g_{id}}f_Z(z_{id}|\boldsymbol{\theta})f_{G|Z}(g_{id}|z_{id},\boldsymbol{\gamma})f_{Z^\star,H|Z,G}(z^\star_{id},h_{id}|z_{id},g_{id})\mathrm{d}z_{id}
\end{equation}
The integration with respect to $z_{id}$ might be non-trivial due to the presence of the latent variable $Z$ in both $f_Z$ and $f_{G|Z}$. To make this model easily estimable, we opt to use a discretized version of the probabilities in \eqref{eq:probsG} when defining $f_{G|Z}(g_{id}|z_{id},\boldsymbol{\gamma})$. In particular, we re-define $G$ as a categorical random variable with probabilities
\begin{equation}\label{eq:discr_prob}
\mathbb{P}[G=g|z,\boldsymbol{\gamma}]=\tilde{\lambda}_g(z,\boldsymbol{\gamma});\quad \tilde{\lambda}_g(z,\boldsymbol{\gamma})= \sum_{q=0}^{+\infty}\lambda_g(q,\boldsymbol{\gamma})\boldsymbol{1}_{I_1(q)}(z),\quad g\in\mathcal{G}.
\end{equation}
By plugging \eqref{eq:discr_prob} and \eqref{eq:con_joint} into the likelihood \eqref{eq:lik_init}, and exploiting the linearity of the integral, we obtain a computationally feasible expression for the likelihood:
\begin{equation}\label{eq:lik_real}
    f_{Z^\star,H}(z^\star_{id},h_{id}|\boldsymbol{\theta,\boldsymbol{\gamma}})=\sum_{g_{id}\in h_{id}}\sum_{q\in\mathcal{C}_{g_{id}}(z^\star_{id})}{\lambda}_{g_{id}}(q,\boldsymbol{\gamma})\int_{I_{1}(q)}f_Z(z_{id}|\boldsymbol{\theta})\mathrm{d}z_{id}
\end{equation}
where $\mathcal{C}_g(z^\star) = \mathbb{N}\cap I_{g}(z^\star)$. Note that the integral that remains in the likelihood of Equation~\eqref{eq:lik_real} can be expressed in terms of the CDF of the distribution of $Z$ as follows
$$
\int_{I_{1}(q)}f_Z(z_{id}|\boldsymbol{\theta})\mathrm{d}z_{id} = \mathbb{P}\left[Z\leq q+\frac{1}{2}\bigg|\boldsymbol{\theta}\right]-\mathbb{P}\left[Z\leq q-\frac{1}{2}\bigg|\boldsymbol{\theta}\right],
$$
making it easier to use this likelihood in practice.

A final remark on the likelihood structure is needed to account for the presence of values censored at 21. Indeed, this value has been assigned to all answers with values greater than or equal to 21. It is worth highlighting that this feature interacts with the latent structure for coarsening behavior. In particular, observing $z^\star=21$ has different implications for $Z$ according to the different regimes of $G$:
$$
\begin{cases}
    Z\geq 20.5 \text{ if } g=1;\\
    Z\geq 22.5 \text{ if } g=5;\\
    Z\geq 24.5 \text{ if } g=10.
\end{cases}
$$
This translates into a likelihood contribution that is independent of $H$ and is equal to:
\begin{align*}
 f_{Z^\star}(21|\boldsymbol{\theta,\boldsymbol{\gamma}})=&\sum_{q\in\{21,22,23,24\}}\lambda_1(q,\boldsymbol{\gamma})\int_{I_{1}(q)}f_Z(z_{id}|\boldsymbol{\theta})\mathrm{d}z_{id} +\\
 &\quad \sum_{q\in\{23,24\}}\lambda_5(q,\boldsymbol{\gamma})\int_{I_{1}(q)}f_Z(z_{id}|\boldsymbol{\theta})\mathrm{d}z_{id}+\\
 &\quad \int_{24.5}^{+\infty}f_Z(z_{id}|\boldsymbol{\theta})\mathrm{d}z_{id}.
\end{align*}

To specify the model likelihood, a distribution for the latent variable $Z$ is required. To enhance flexibility, we adopt a finite mixture model of Lognormal distributions, which has already been proposed for unit-level small area modeling in \citet{gardini2022poverty}. In this work, the proposed model is slightly different since, in addition to component-specific scale parameters, we include component-specific intercepts and allow the probability of belonging to a given component to depend on a set of covariates, entering the class of so-called Mixture of Experts models \citep{celeux2018handbook}. Given the restricted support of the target response variable, we consider here a model with two mixture components, although the specification can easily be extended to include more components. The distribution is defined as
\begin{equation}
    Z_{id}|l_{id},\boldsymbol{\theta}\sim \boldsymbol{1}_{\{1\}}(l_{id})\mathcal{LN}\left(\mu_{1,id},\sigma^2_1\right)+\boldsymbol{1}_{\{2\}}(l_{id})\mathcal{LN}\left(\mu_{2,id},\sigma^2_2\right);
\end{equation}
where $l_{id}\in\{1,2\}$ is the latent label determining the component assigned to each unit and 
$
\mu_{l_{id},id}=\beta_{0l_{id}}^\mu+\mathbf{x}_{id}^\top\boldsymbol{\beta}^\mu+u^\mu_{d}
$
is the location parameter of component $l_{id}$.
The probability of the latent label is crucial in a mixture model and we let it depend on covariates as follows:
\begin{equation}
    \mathbb{P}\left[l_{id}=1|\mathbf{x}_{id}\right]=\pi_{id}=\text{expit}\left(\beta_0^\pi+\mathbf{x}_{id}^\top\boldsymbol{\beta}^\pi+u_d^\pi\right).
\end{equation}
 In both predictors, we adopt a standard mixed-effects specification: $\beta_0^\pi$ and $\beta_{0l}^\mu$ denote the intercepts, $\boldsymbol{\beta}^\pi$ and $\boldsymbol{\beta}^\mu$ denote the regression coefficients, while $u_d^\pi$ and $u_d^\mu$ are domain-specific random effects. As in the first part of Section \ref{sec:part1}, we assume Gaussian priors for these random effects, namely $u_d^\pi|\tau_\pi \stackrel{ind}{\sim}\mathcal{N}(0,\tau_\pi^2)$ and $u_d^\mu|\tau_\mu \stackrel{ind}{\sim}\mathcal{N}(0,\tau_\mu^2)$, $\forall d$. For the following developments, we define the subset of model parameters governing the mixing probability as $\boldsymbol{\theta}_\pi=(\beta_0^\pi,\boldsymbol{\beta}^\pi,\mathbf{u}^\pi)\subset\boldsymbol{\theta}$.

Now, it is possible to express the mixture distribution by marginalizing with respect to the latent label, obtaining the density function for the Lognormal Mixture (LNM) model:
\begin{equation}\label{eq:distr_LNM}
    f_Z^\text{LNM}(z_{id}|\boldsymbol{\theta})= \pi_{id}f_Z^{LN}\left(z_{id}|\mu_{1,id},\sigma^2_1\right)+\left[1-\pi_{id}\right]f_Z^{LN}\left(z_{id}|\mu_{2,id},\sigma^2_2\right),
\end{equation}
where $f_Z^{LN}\left(z|\mu,\sigma^2\right)$ is the density function of a Lognormal random variable evaluated at $z$ with parameters $\mu$ and $\sigma^2$. For the sake of comparison, in the next sections of the paper we also consider the simple Lognormal (LN) model, defined as
\begin{equation}
    Z_{id}|\boldsymbol{\theta}\sim \mathcal{LN}\left(\beta_{0}^\mu+\mathbf{x}_{id}^\top\boldsymbol{\beta}^\mu+u^\mu_{d},\sigma^2\right).
\end{equation}

To finalize the model specification, prior distributions for the remaining parameters are needed, namely for the regression coefficients and the variance components; this topic will be discussed in detail in Section \ref{sec:prior}.

\subsection{Model fitting, posterior inference and model checking}\label{sec:p3}

The population two-part model introduced in Sections \ref{sec:part1} and \ref{sec:part2} is fitted using the sampled units. To this end, let $\mathbf{w}_s$, $\mathbf{z}_s^\star$, and $\mathbf{h}_s$ denote the $n$-dimensional vectors collecting the corresponding sample values. We recall that the participation and intensity components are conditionally independent; hence, the likelihood factorizes and the two components can be estimated separately. For the continuous component, we obtain draws from the joint posterior $p(\boldsymbol{\theta},\boldsymbol{\gamma}| \mathbf{z}_s^\star,\mathbf{h}_s)$ and, whenever convenient, we work with the induced marginal posteriors $p(\boldsymbol{\theta}| \mathbf{z}_s^\star,\mathbf{h}_s)$ and $p(\boldsymbol{\gamma}| \mathbf{z}_s^\star,\mathbf{h}_s)$.

The proposed models were fit using Hamiltonian Monte Carlo, a Markov Chain Monte Carlo (MCMC) method readily available in the \texttt{Stan} programming language and accessed through its \texttt{rstan} interface \citep{rstan}. A multi-chain approach was adopted by running 4 parallel chains with 2,000 iterations each, discarding the first 1,000 iterations as the warm-up period, thus obtaining $B=4,000$ draws from the posterior distributions of the model parameters $p(\boldsymbol{\delta}|\mathbf{w}_s)$, $p(\boldsymbol{\theta}|\mathbf{z}^\star_s,\mathbf{h}_s)$, and $p(\boldsymbol{\gamma}|\mathbf{z}^\star_s,\mathbf{h}_s)$. The notation $\boldsymbol{\delta}^{(b)}$ is used to indicate the generic $b$-th draw from the posterior distribution of the considered parameter.

While fitting the logistic regression related to the variable $W$ represents a standard task, some ad hoc expedients are necessary to fit the model for $Z$. Indeed, for identifying  model components we assume $\beta_{01}^\mu<\beta_{02}^\mu$. It is worth noting that component identification is not central to our aims, as our primary objective is to predict the response, and predictive performance does not depend on the specific labeling of the mixture components.

Having obtained the MCMC draws from the posterior distribution, inference on the target parameters and on additional quantities of interest becomes straightforward. To perform inference on the target indicators at the domain level, we require the posterior predictive distributions of the dichotomous variable $W$ and the latent variable $Z$. The first is the standard posterior predictive distribution, and a generic unobserved unit $\widetilde{w}$ has distribution
\begin{equation}\label{eq:postpred_w}
    p(\widetilde{w}|\mathbf{w}_s)=\int_\Delta f_W(\widetilde{w}|\boldsymbol{\delta})p(\boldsymbol{\delta}|\mathbf{w}_s)\mathrm{d}\boldsymbol{\delta}.
\end{equation}
Similarly, it is possible to define also the distribution of an unobserved unit $\tilde{z}$ of the latent variable $Z$:
\begin{equation}\label{eq:postpred_z}
p(\tilde{z}|\mathbf{z}^\star_s,\mathbf{h}_s)=\int_\Theta f_Z(\tilde{z}|\boldsymbol{\theta})p(\boldsymbol{\theta}|\mathbf{z}^\star_s,\mathbf{h}_s)\mathrm{d}\boldsymbol{\theta}.
\end{equation}
These integrals can be numerically solved thanks to the draws from the posterior distributions. Specifically, $\widetilde{w}^{(b)}$ and $\tilde{z}^{(b)}$ can be generated from the distribution of $W$ and $Z$ by setting the model parameters equal to $\boldsymbol{\delta}^{(b)}$ and $\boldsymbol{\theta}^{(b)}$, respectively. 

Subsequently, it is possible to define the posterior predictive distribution for the latent variable $G$ to predict the coarsening level:
\begin{equation}
p(\tilde{g}|\mathbf{z}^\star_s,\mathbf{h}_s)=\int_\Gamma\int_{\mathbb{R}^+} f_{G|Z}(\tilde{g}|\tilde{z},\boldsymbol{\gamma})p(\tilde{z}|\mathbf{z}^\star_s,\mathbf{h}_s)p(\boldsymbol{\gamma}|\mathbf{z}^\star_s,\mathbf{h}_s)\mathrm{d}\tilde{z}\mathrm{d}\boldsymbol{\gamma}.
\end{equation}
Also in this case, a draw of $\tilde{g}^{(b)}$ can be obtained by combining the $b$-th draw from the posterior predictive distribution of $\tilde{z}$ and $\boldsymbol{\gamma}^{(b)}$. Having both $\tilde{z}^{(b)}$ and $\tilde{g}^{(b)}$ allows us to obtain posterior draws for the quantity $\tilde{z}^\star$, i.e. the coarsened data:
$$
\tilde{z}^{\star(b)} = T_{Z^\star}(\tilde{z}^{(b)},\tilde{g}^{(b)}).
$$
The distribution $\tilde{z}^\star|\mathbf{z}^\star_s,\mathbf{h}_s$ can be particularly useful to check the performance of the model. Indeed, posterior predictive checks can be performed by comparing the data variables generated under the model with the observed vector $\mathbf{z}_s^\star$ \citep{gelman2005multiple}.

\subsection{Hierarchical Bayes predictors of target quantities}\label{sec:p4}

The posterior predictive distributions defined in equations \eqref{eq:postpred_w} and \eqref{eq:postpred_z} can be exploited to predict the domain-specific target quantities defined at the population level in Section \ref{sec:target}, thus obtaining their posterior distributions under the small area model. 

In practice, the $b$-th draw from the posterior distribution of $\overline{w}_d$ is defined by combining the sample information in $\mathbf{w}_{s,d}=(w_{1d},\dots,w_{n_dd})$ and the predictions of the outcomes for the $N_d-n_d$ unsampled units. The predictions are obtained by exploiting the available auxiliary information, and the $b$-th set of values drawn from the posterior predictive distribution is denoted as $\tilde{\mathbf{w}}_{\bar{s},d}^{(b)}=\left(\widetilde{w}^{(b)}_{(n_d+1)d},\dots,\widetilde{w}^{(b)}_{N_dd}\right)$. The final draw from the distribution $\overline{w}_d|\mathbf{w}_s$ is obtained as
\begin{equation*}
    \overline{w}_d^{(b)}|\mathbf{w}_s=\frac{1}{N_d}\left(\sum_{i=1}^{n_d}w_{id}+\sum_{i=n_d+1}^{N_d}\widetilde{w}^{(b)}_{id}\right),\quad \forall d.
\end{equation*}

The situation is different for the functionals that involve the average number of cigarettes smoked per day. Indeed, for this quantity we only observe the coarsened and censored value $Z^\star$. For this reason, the predictive distribution \eqref{eq:postpred_z} is also used to generate the values assigned to sampled units, obtaining the following draws from $\bar{z}_d|\mathbf{w}_s,\mathbf{z}_s^\star,\mathbf{h}_s$
\begin{equation*}
\bar{z}_d^{(b)}|\mathbf{w}_s,\mathbf{z}_s^\star,\mathbf{h}_s=\frac{\sum_{i=1}^{n_d}w_{id}\tilde{z}_{id}^{(b)}+\sum_{i=n_d+1}^{N_d}\widetilde{w}^{(b)}_{id}\tilde{z}_{id}^{(b)}}{\sum_{i=1}^{n_d}w_{id}+\sum_{i=n_d+1}^{N_d}\widetilde{w}^{(b)}_{id}},\quad \forall d;
\end{equation*}
and from $\overline{HS}|\mathbf{w}_s,\mathbf{z}_s^\star,\mathbf{h}_s$
\begin{equation*}
\overline{HS}_d^{(b)}|\mathbf{w}_s,\mathbf{z}_s^\star,\mathbf{h}_s=\frac{\sum_{i=1}^{n_d}\mathbf{1}_{[20,\infty)}\left(w_{id}\tilde{z}_{id}^{(b)}\right)+\sum_{i=n_d+1}^{N_d}\mathbf{1}_{[20,\infty)}\left(\widetilde{w}^{(b)}_{id}\tilde{z}_{id}^{(b)}\right)}{\sum_{i=1}^{n_d}w_{id}+\sum_{i=n_d+1}^{N_d}\widetilde{w}^{(b)}_{id}},\quad \forall d.
\end{equation*}

Once a sample from the posterior distribution is available, we propose as a point estimator the posterior mean, which is the optimal Bayes estimator under quadratic loss and can be easily approximated through Monte Carlo methods:
$$
\hat{\zeta}_d = \mathbb{E}\left[\zeta_d|\mathbf{w}_s,\mathbf{z}_s^\star,\mathbf{h}_s\right]\approx\frac{1}{B}\sum_{b=1}^B\zeta_d^{(b)},
$$
where $\zeta_d\in\{\overline{w}_d,\bar{z}_d,\overline{HS}_d\}$.

\subsection{Prior distributions}\label{sec:prior}

The Bayesian models defined in Sections \ref{sec:part1} and \ref{sec:part2} require setting prior distributions for the parameters. We use weakly informative priors for all the regression coefficients, basically following the indications in \citet{gelman2017prior}. In particular, we set a $\mathcal{N}(0,2.5^2)$ distribution for the intercepts $\beta_0^\nu$, $\beta_0^\pi$, $\gamma_{01}$, and $\gamma_{02}$; whereas a $\mathcal{N}(m_{\log(\mathbf{z}^\star)},2.5^2s_{\log(\mathbf{z}^\star)}^2)$ is assumed for $\beta_{01}^\mu$ and $\beta_{02}^\mu$, defining $m_{\log(\mathbf{z}^\star)}$ and $s_{\log(\mathbf{z}^\star)}$ as the mean and the standard deviation of $\log(\mathbf{z}^\star)$, respectively. Considering the regression coefficients, we remark that all the covariates are centered at 0 with unit variance, and independent $\mathcal{N}(0,2.5^2)$ priors are set for the elements of $\boldsymbol{\beta}^\nu$ and $\boldsymbol{\beta}^\pi$, along with $\gamma_1$. For the elements of $\boldsymbol{\beta}^\mu$, independent $\mathcal{N}(0,2.5^2s_{\log(\mathbf{z}^\star)}^2)$ priors are assumed.

More attention is required when specifying priors for the scale parameters $\sigma_1$, $\sigma_2$, $\tau_\nu$, $\tau_\pi$, and $\tau_\mu$. Indeed, these choices can affect the existence of posterior moments for relevant quantities under the LN and LNM specifications, as discussed in \citet{gardini2022poverty}. We first state a condition ensuring the existence of moments for the posterior predictive distribution of $Z$.

\begin{prop}\label{prop:moments_pred}
Let us consider the likelihood accounting for heaping in \eqref{eq:lik_real} and the LNM distributional assumption in \eqref{eq:distr_LNM}. Gaussian priors with known parameters are set for $\beta_{01}^\mu$, $\beta_{02}^\mu$, and $\boldsymbol{\beta}^\mu$. Moreover, the random effects are assigned a Gaussian prior, $u_{d}^\mu|\tau_\mu\sim\mathcal{N}(0,\tau^2_\mu)$, and proper priors are set for the remaining parameters. Under this modeling framework, the $r$-th posterior predictive moment $\mathbb{E}\left[\tilde{z}^r_{id}|\mathbf{z}_s^\star,\mathbf{h}_s\right]$, $\forall i,d$, is finite if and only if the priors for $\sigma_1$, $\sigma_2$, and $\tau_\mu$ have densities with an exponential term that goes to 0 faster than
\begin{equation}\label{eq:conditions}
\exp\left\{-\frac{r^2}{2}\sigma_1^2\right\},\quad \exp\left\{-\frac{r^2}{2}\sigma_2^2\right\},\quad \exp\left\{-\frac{r^2}{2}\tau_\mu^2\right\}.        
\end{equation}
\end{prop}
\begin{proof}
See Supplementary Material.
\end{proof}

To guide the choice of priors for the scale parameters, we next state a proposition providing existence conditions for the posterior moments of the target indicators $\bar{z}_d$ and $\overline{HS}_d$.

\begin{prop}\label{prop:moments}
Let us consider the likelihood accounting for heaping in \eqref{eq:lik_real} and the LNM distributional assumption in \eqref{eq:distr_LNM}. Gaussian priors with known parameters are set for $\beta_{01}^\mu$, $\beta_{02}^\mu$ and $\boldsymbol{\beta}^\mu$. Moreover, the random effects are assigned a Gaussian prior, $u_{d}^\mu|\tau_\mu\sim\mathcal{N}(0,\tau^2_\mu)$, and proper priors are set for the remaining parameters. Under this modeling framework, assume that
\begin{equation}\label{eq:condition_W}
\exists \varepsilon>0\text{ s.t. }\mathbb{P}\left[\sum_{i=1}^{n_d}w_{id}+\sum_{i=n_d+1}^{N_d}\widetilde{w}_{id}>\varepsilon|\mathbf{w}_s\right]=1;
\end{equation}
Then, the following existence conditions hold:
\begin{enumerate}
    \item $\mathbb{E}\left[\overline{HS}_d^r|\mathbf{w}_s,\mathbf{z}_s^\star,\mathbf{h}_s\right]<+\infty$ under any proper prior choice for $\sigma_1$, $\sigma_2$, and $\tau_\mu$.
    \item $\mathbb{E}\left[\bar{z}_d^r|\mathbf{w}_s,\mathbf{z}_s^\star,\mathbf{h}_s\right]<+\infty$ if and only if the conditions listed in \eqref{eq:conditions} that guarantee the existence of $\mathbb{E}\left[\tilde{z}_{id}^r|\mathbf{z}_s^\star,\mathbf{h}_s\right]$ hold.
\end{enumerate}
\end{prop}
\begin{proof}
See Supplementary Material.
\end{proof}
We remark that the assumption in \eqref{eq:condition_W} is not restrictive in our applied setting, as it essentially means that at least one daily smoker is observed in each subpopulation, and this is certainly satisfied since $\sum_{i=1}^{n_d}w_{id}\geq1,\ \forall d$. We also note that, although the statement and proof are given for the LNM specification, the same conditions carry over to the degenerate LN case.

Keeping the idea of specifying weakly informative prior distributions, we opt for a Half-Normal (HN) prior with scale parameter equal to 2 for the scale parameters not involved in Propositions \ref{prop:moments_pred} and \ref{prop:moments}, namely $\tau_\nu$ and $\tau_\pi$. To guarantee the existence of posterior moments for the target quantity $\bar{z}_d$, we consider the Generalized Half-Normal (GHN) distribution \citep{cooray2008generalization}. If $X\sim\text{GHN}(a,b)$, then its density is
$$
f(x|a,b)=\sqrt{\frac{2}{\pi}}\left(\frac{a}{x}\right)\left(\frac{x}{b}\right)^a\exp\left\{-\frac{1}{2}\left(\frac{x}{b}\right)^{2a}\right\}.
$$
By setting $a=3/2$, we obtain a prior that guarantees the condition in Proposition \ref{prop:moments}. The scale parameter is set to $b=2.788$ in order to match the variance of the HN prior with scale equal to 2.

\section{Simulation study}\label{sec:4}

In this section, we present a simulation study aimed at quantifying the consequences of ignoring coarsening when it is present, and at comparing the proposed model with competing approaches for estimating $\bar{z}_d$ and $\overline{HS}_d$. The estimation of $\overline{w}_d$ is not considered, as this problem has already been extensively examined in the literature \citep[see, among others,][]{dreassi2014small, chandra2016small}. The simulation considers several heaping scenarios that differ in the available rounding levels and in whether the heaping mechanism is informative with respect to the latent intensity $Z$.

In the simulation study, the synthetic population is generated under a two-component Lognormal mixture, with parameters varying across areas and depending on a covariate. All parameters are calibrated to closely match those observed in the empirical application, in order to mimic a realistic data-generating process. The population comprises $N=\sum_{d=1}^{30} N_d$ units, with $N_d$ taking values 700, 1{,}000, and 1{,}300 for ten areas each. 
For each unit, a binary covariate $x_{id}$ is first generated with $\mathbb{P}[x_{id}=1]=0.4$. Conditional on $x_{id}$, the mixing probability is defined as $p_{id}=\text{expit}(0.4+0.2x_{id})$, and the component-specific location parameters are specified as
$m_{l_{id},id}=b_{0l_{id}}^\mu+0.05x_{id}+u_d^\mu$,
with $b_{01}^\mu=1.7$, $b_{02}^\mu=2.7$, and $u_d^\mu\sim\mathcal{N}(0,0.25^2)$, drawn once per area. Each unit is then assigned to component $l_{id}\in\{1,2\}$ according to $p_{id}$, and the latent intensity $Z$ is generated from $\mathcal{LN}(m_{1,id},0.5^2)$ if $l_{id}=1$, or from $\mathcal{LN}(m_{2,id},0.25^2)$ if $l_{id}=2$.

From the population, we extract $R=500$ samples by stratified simple random sampling without replacement, using a sampling fraction equal to $3\%$. For each extracted sample, we keep the same sampled units and the same underlying latent values of $Z$, and we only modify the way the observed answer $Z^\star$ is generated, so as to isolate the effect of the heaping mechanism. Specifically, for each sampled unit we first generate a latent heaping indicator $G$ conditional on the latent intensity $Z$, and then we obtain the reported value $Z^\star$ by applying the coarsening rule encoded in the transformation $T_{Z^\star}(Z,G)$ introduced in Section~\ref{sec:part2}. When the resulting value exceeds 20, we apply the same top-coding rule used in the survey and record $Z^\star=21$.

Across scenarios, the heaping mechanism is generated through the probabilities defined in \eqref{eq:probsG}. The parameter $\gamma_1$ determines whether coarsening is ignorable ($\gamma_1=0$) or depends on $Z$ ($\gamma_1\neq 0$). 
The first two scenarios consider heaping only at multiples of 5, so that $G$ can take only two values (rounding to the nearest integer or to the nearest multiple of 5). In this case, a reduced version of \eqref{eq:probsG}, governed by $(\gamma_{0},\gamma_{1})$, is used.
\begin{itemize}
    \item \textbf{Scenario 1}. Heaping at multiples of 5 without dependence on $Z$. The parameters are set to $\gamma_{0} = 2.0$ and $\gamma_{1} = 0$. 
    \item \textbf{Scenario 2}. Heaping at multiples of 5 with dependence on $Z$. The parameters are set to $\gamma_{0} = 5.5$ and $\gamma_{1} = -3.2$.
\end{itemize}
In the last two scenarios, the heaping probability follows the full specification in \eqref{eq:probsG}, allowing for rounding at multiples of both 5 and 10.
\begin{itemize}
    \item \textbf{Scenario 3}. Heaping at multiples of 5 and 10 without dependence on $Z$. The parameters are set to $\gamma_{01} = 0.5$, $\gamma_{02} = 2.5$ and $\gamma_{1} = 0$.
    \item \textbf{Scenario 4}. Heaping at multiples of 5 and 10 with dependence on $Z$. The parameters are set to $\gamma_{01} = 7.0$, $\gamma_{02} = 9.7$ and $\gamma_{1} = -3.4$.
\end{itemize}

For each replication $r = 1, \dots, R$, $\widehat{\zeta}^{(r)}_d$ is computed for $\zeta_d \in \{\bar{z}_d, \overline{HS}_d\}$ using four estimators based on: a Lognormal model without coarsening (LN), its extension including coarsening (LN-C), a two-component Lognormal mixture model (LNM), and the corresponding version accounting for coarsening (LNM-C). Performance is evaluated, for a generic domain $d$, through the relative bias (RB) and the relative root mean square error (RRMSE):
$$
\text{RB}(\hat{\zeta}_d) =  \frac{1}{R} \sum_{r=1}^R \left(\frac{\hat{\zeta}_{d}^{(r)}}{\zeta_d} -1\right),\quad \text{RRMSE} (\hat{\zeta}_d) =  \sqrt{\frac{1}{R} \sum_{r=1}^R 
    \frac{(\hat{\zeta}_{d}^{(r)} - \zeta_d)^2}{\zeta_d^2}}.
$$
Frequentist properties of the credible intervals are assessed by defining $\hat{\zeta}^{(r)}_{d,\inf}$ and $\hat{\zeta}^{(r)}_{d,\sup}$ as the lower and upper bounds of the $90\%$ credible interval for domain $d$ at replication $r$, and computing the frequentist coverage (Cov) and the interval width (W):
$$
\text{Cov}(\hat{\zeta}_d)= \frac{1}{R} \sum_{r=1}^R 
    \boldsymbol{1}_{\left[\hat{\zeta}^{(r)}_{d,\inf},\,  \hat{\zeta}^{(r)}_{d,\sup}\right]}( \zeta_d ),\quad  \text{W}(\hat{\zeta}_d) =   \frac{1}{R} \sum_{r=1}^R  \left(\hat{\zeta}^{(r)}_{d,\sup} - \hat{\zeta}^{(r)}_{d,\inf} \right).
$$
Overall performance measures are obtained by averaging these indicators over the $D$ domains, defining
$\text{AI}=\frac{1}{D}\sum_{d=1}^D\text{I}(\hat{\zeta}_d)$, with $\text{I}\in\{\text{RB}, \text{RRMSE}, \text{Cov}, \text{W}\}$.

\begin{table}[]
\caption{Average measures of performance for model-based estimators of $\overline{z}_d$ and $\overline{HS}_d$}
\label{tab:1}
\centering
\begin{tabular}{@{}llrrrrrrrr@{}}
\toprule
                                                     &       & \multicolumn{4}{c}{Estimators of $\bar{z}_d$}                              & \multicolumn{4}{c}{Estimators of $\overline{HS}_d$}                                         \\ \cmidrule(l){3-6} \cmidrule(l){7-10}
                                        & Model & ARB                   & ARRMSE                & ACov     & AW   & ARB              & ARRMSE           & ACov & AW \\ \midrule
\multirow{4}{*}{\rotatebox{90}{Scenario 1}} & LN    & 0.093 & 0.140 & 0.841 & 4.782 & 1.775 & 1.931 & 0.434 & 0.138 \\ 
                          & LN-C  & 0.088 & 0.136 & 0.852 & 4.745 & 1.726 & 1.885 & 0.442 & 0.137 \\ 
                           & LNM   & 0.013 & 0.073 & 0.903 & 2.917 & 0.271 & 0.582 & 0.896 & 0.130 \\ 
                           & LNM-C & 0.008 & 0.071 & 0.901 & 2.859 & 0.200 & 0.529 & 0.904 & 0.128 \\  \midrule
\multirow{4}{*}{\rotatebox{90}{Scenario 2}}                   & LN    &0.062 & 0.119 & 0.879 & 4.486 & 1.599 & 1.780 & 0.449 & 0.130 \\ 
                                                & LN-C  & 0.064 & 0.119 & 0.885 & 4.546 & 1.538 & 1.718 & 0.467 & 0.133 \\ 
                                                & LNM   & 0.001 & 0.079 & 0.831 & 2.617 & 0.211 & 0.619 & 0.838 & 0.117 \\ 
                                                & LNM-C & 0.008 & 0.076 & 0.897 & 2.941 & 0.213 & 0.567 & 0.898 & 0.130 \\  \midrule
\multirow{4}{*}{\rotatebox{90}{Scenario 3}}                   & LN    & 0.086 & 0.133 & 0.847 & 4.647 & 1.712 & 1.874 & 0.441 & 0.136 \\ 
                                                & LN-C  & 0.084 & 0.132 & 0.858 & 4.679 & 1.670 & 1.834 & 0.450 & 0.137 \\ 
                                                & LNM   & 0.015 & 0.074 & 0.901 & 2.964 & 0.288 & 0.602 & 0.890 & 0.130 \\ 
                                                & LNM-C & 0.013 & 0.073 & 0.905 & 2.989 & 0.250 & 0.573 & 0.901 & 0.132 \\ \midrule
\multirow{4}{*}{\rotatebox{90}{Scenario 4}}                    & LN    & 0.048 & 0.111 & 0.893 & 4.354 & 1.492 & 1.688 & 0.455 & 0.127 \\ 
                                                & LN-C  & 0.043 & 0.108 & 0.905 & 4.377 & 1.414 & 1.617 & 0.477 & 0.128 \\ 
                                                & LNM   & 0.019 & 0.133 & 0.528 & 1.954 & 0.841 & 1.602 & 0.575 & 0.103 \\ 
                                                & LNM-C & -0.003 & 0.074 & 0.896 & 2.971 & 0.128 & 0.526 & 0.901 & 0.130 \\   \bottomrule
\end{tabular}
\end{table}

Table~\ref{tab:1} reports domain-averaged performance measures across the four scenarios for both target indicators. 
In Scenarios~1 and~3, where the coarsening mechanism does not depend on $Z$, coarsened and non-coarsened versions of the same model exhibit very similar performance. Scenario~2 shows improved results for the coarsening-aware specifications (LN-C and LNM-C), particularly in terms of RRMSE and coverage. A different pattern emerges in Scenario~4: when coarsening affects both multiples of 5 and 10 and depends on $Z$, the LNM estimator without coarsening correction displays marked coverage deterioration and instability, whereas the LNM-C specification remains robust and restores near-nominal coverage for both target parameters. This contrast is particularly evident for $\overline{HS}_d$, where coverage drops below 0.60 for LNM while remaining close to 0.90 under the LNM-C specification.

\begin{figure}
    \centering
    \includegraphics[width=1\linewidth]{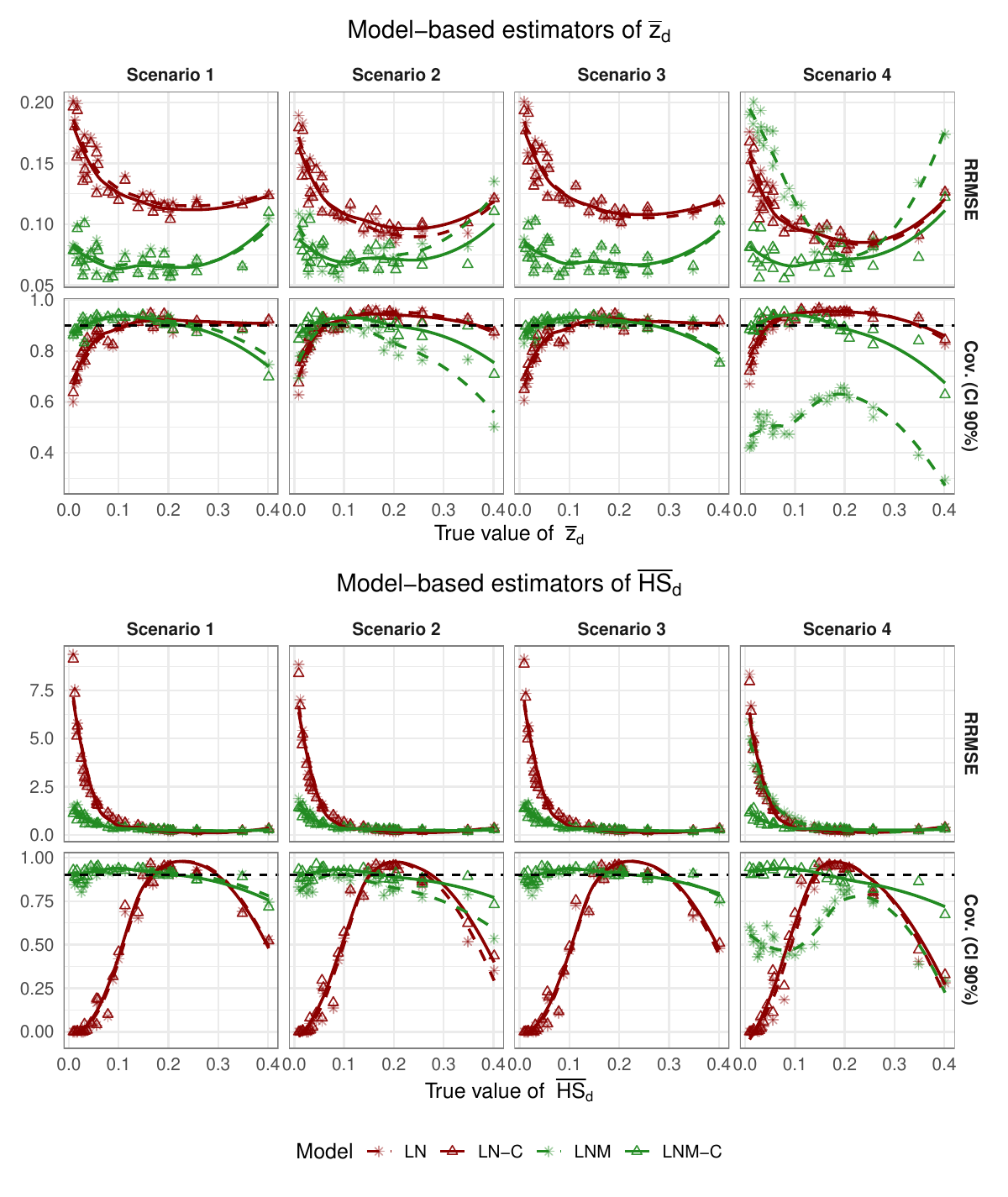}
    \caption{Domain-specific measures of performance for model-based estimators of $\overline{z}_d$ and $\overline{HS}_d$}
    \label{fig:2}
\end{figure}

Across scenarios, and especially when coarsening depends on $Z$ (Scenarios~2 and~4), accounting for both distributional heterogeneity and coarsening is crucial. When the heaping mechanism is properly modeled, estimators based on the LNM-C model outperform their LN-C counterparts by reducing bias and RRMSE and achieving coverage probabilities closer to the nominal level, often with comparable or shorter average interval widths. For $\bar{z}_d$, improvements are more moderate, reflecting the lower sensitivity of mean-type indicators to distributional misspecification. By contrast, $\overline{HS}_d$ is intrinsically more fragile: being threshold-based, it is highly sensitive to how probability mass is allocated around the cutoff at 20 cigarettes, so even small distortions induced by coarsening can translate into substantial bias and undercoverage. This is most evident in Scenario~4, where failing to model both mixture features and heaping leads to severe undercoverage for LNM, while the LNM-C specification attains near-nominal coverage and markedly improved accuracy (ARB 0.128 versus values above 0.8 for all other specifications; ARRMSE 0.526 versus values above 1.6).

Figure~\ref{fig:2} and Figure~S1 in the Supplementary Material report RRMSE, Cov, RB, and W, respectively. Unlike Table~\ref{tab:1}, these figures display performance measures at the domain level as a function of the true value of the estimands, which helps clarify how performance varies across the support. The patterns observed for the averaged indicators are substantially confirmed, while the domain-level view provides additional insight into where differences between models arise.
For $\bar{z}_d$, the RRMSE exhibits a clear convex pattern with respect to the true value, with lower errors for domains whose $\bar{z}_d$ is close to the overall mean. This dependence is largely reduced by mixture models in the first three scenarios, while in Scenario~4 a pronounced concave-shape persists for LN, LN-C, and LNM and is effectively removed by LNM-C, which yields a nearly flat RRMSE profile across the support. Coverage shows the complementary behavior, being more stable and closer to the nominal level under LNM-C. For $\overline{HS}_d$, the same qualitative conclusions hold: compared to $\bar{z}_d$, the convex pattern is less pronounced and larger errors mainly occur for low values of $\overline{HS}_d$; mixture-based models outperform single-distribution specifications in the first three scenarios, and in Scenario~4 the advantage of LNM-C becomes particularly evident, delivering the lowest RRMSE across the support and coverage patterns consistent with those observed for $\bar{z}_d$.

To conclude, the Supplementary Material also reports the performance measures for the design-based estimators (Table~S1). Specifically, the direct estimator computed on the true sample (i.e., without coarsening) is compared with its counterpart obtained under the four simulated coarsening scenarios, for both target indices. As expected, in the absence of coarsening the direct estimator is essentially unbiased for both quantities. In the presence of coarsening, however, especially for $\overline{HS}_d$, bias increases substantially and precision deteriorates, making the model-based estimators preferable.

\section{Smoking behaviour in Italy}\label{sec:5}

This section applies the unit-level Bayesian SAE models introduced in Section~\ref{sec:3} to the Italian EHIS data described in Section~\ref{sec:2}, with the aim of producing small area estimates of smoking indicators for region--age domains while comparing alternative model specifications.

The proposed approach follows a two-part framework: computation for the dichotomous component $W$ is fast, whereas fitting the coarsening-aware models for $Z^\star$ is more demanding. All models were fitted on a shared server equipped with two AMD EPYC 7763 processors (64 cores each, 2.45~GHz), for a total of 128 cores, and 2~TB of RAM; under this setup, the LN-C model required approximately 130 minutes of computation, whereas the LNM-C model required about 360 minutes.

In what follows, Section~\ref{sec:5.1} compares results across the four models considered in the simulation study, after which Section~\ref{sec:5.2} presents SAE estimates and discusses their implications for targeted public health policies. 

\subsection{Model Selection and Validation}\label{sec:5.1}

\begin{figure}
    \centering
\includegraphics[width=.9\linewidth]{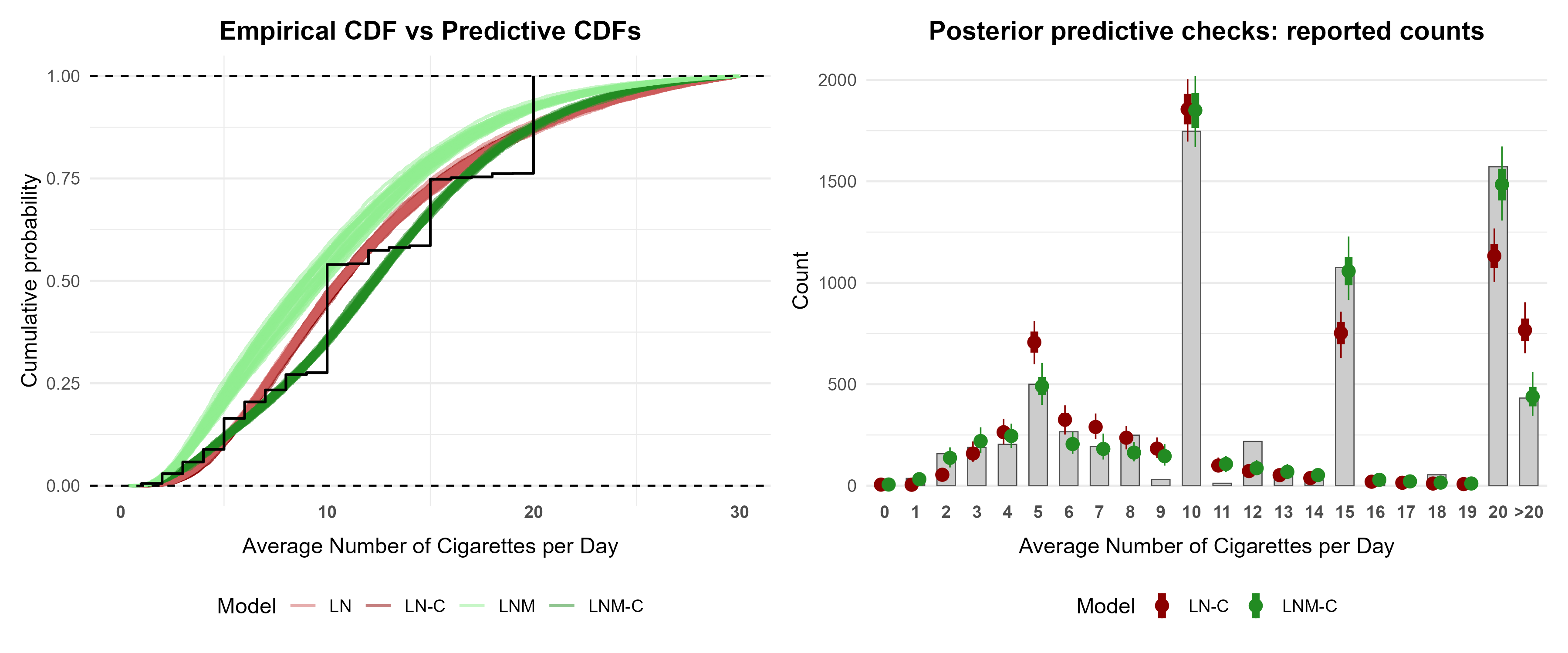}
    \caption{Posterior predictive checks. Left-plot: Simulated predictive CDFs of $Z$ under different models, compared with the empirical CDF of $Z^\star$ (black line). Right-plot: Frequency distribution of $Z^\star$ (gray bars) compared with the distribution of predicted counts under LN-C and LNM-C models.}
        \label{fig:app1}
\end{figure}

A first assessment of the fitted models is performed via posterior predictive checks, which compare features of the observed data with those of replicated data generated from the posterior predictive distribution. Focusing on the binary component for daily smoking, the sample proportion is 0.154, which falls within the 90\% posterior predictive credible interval $[0.150,\,0.158]$, thus showing no relevant lack of fit.

The focus then moves to the second component, where both the outcome distribution and the specification of the coarsening mechanism are crucial. Figure~\ref{fig:app1} (left panel) compares the empirical CDF of $Z^\star$ with the posterior predictive CDFs of $Z$ under the four models (LN, LN-C, LNM, and LNM-C). Consistently with the simulation evidence, LN and LN-C are virtually indistinguishable, whereas the difference between LNM and LNM-C is marked. In particular, the LNM model, which ignores coarsening, deviates substantially from the empirical CDF, while the LNM-C specification closely tracks the observed pattern.

To further assess whether the model also captures the coarsening process, the right panel of Figure~\ref{fig:app1} compares the empirical bar plot with posterior predictive counts from the two coarsening-aware models. The LNM-C model reproduces the empirical counts well across the entire distribution, whereas LN-C shows noticeable departures, especially in the right tail: it underestimates the frequency of 20s and overestimates values above 20. Finally, Figure~S2 in the Supplementary Material compares posterior predictive counts under LNM-C when allowing for rounding at multiples of 5 only versus at multiples of both 5 and 10, showing that including the latter substantially improves the fit.

\begin{table}
\centering
\begin{tabular}{lrrrr}
\toprule
 & \multicolumn{2}{c}{LN-C} &  \multicolumn{2}{c}{LNM-C} \\\cmidrule(lr){2-3}\cmidrule(lr){4-5}
Par. & Post. Mean & 90\% C.I. & Post. Mean & 90\% C.I. \\
\midrule
$\gamma_{01}$     & 6.966 & [ 6.509,\,7.421] & 7.010 & [6.541,\  7.498] \\
$\gamma_{02}$     & 9.905 & [9.366,\,10.448] & 9.743 & [9.204,\,10.296] \\
$\gamma_1$         & -3.410 & [ -3.607,\,-3.214] & -3.396 & [ -3.600,\,-3.199] \\
$\beta_{01}^\mu$ 
               &  2.367 & [  2.332,\; 2.400] &  1.972 & [  1.891,\; 2.043] \\
$\beta_{02}^\mu$    &   -  &     -         &  2.633 & [  2.593,\; 2.670] \\
$\tau_\mu$   &  0.168 & [  0.140,\; 0.200] &  0.145 & [  0.107,\; 0.182] \\
$\sigma_1$ 
               &  0.591 & [  0.582,\; 0.600] &  0.681 & [  0.652,\; 0.709] \\
$\sigma_2$     &     - &         -      &  0.313 & [  0.294,\; 0.333] \\
$\beta_0^\pi$   &   -  &      -        & -0.385 & [ -0.639,\,-0.134] \\
$\tau_\pi$ &   - &        -      &  0.503 & [  0.347,\; 0.692] \\\midrule
 & LOOIC & S.E. & LOOIC & S.E. \\\cmidrule(lr){2-3}\cmidrule(lr){4-5}
 &   32427 & 194   &  31441 &  186  \\
\bottomrule
\end{tabular}
\caption{Posterior means and 90\% credible intervals for parameters under the LN-C and LNM-C models.}
\label{tab:LN_LNM}
\end{table}

Inspection of the posterior parameter estimates (Table~\ref{tab:LN_LNM}) supports the use of a mixture specification. Under LNM-C, two components are identified: a lower-consumption component with location $\beta^\mu_{01}=1.972$ and higher dispersion ($\sigma_1=0.681$), and a higher-consumption component with location $\beta^\mu_{02}=2.633$ and lower dispersion ($\sigma_2=0.313$). The corresponding LN-C estimates lie between these two regimes, as expected under a single-component model. This pattern suggests unobserved heterogeneity that is not well captured by a single Lognormal distribution, and the LOOIC comparison is consistent with this, favoring LNM-C over LN-C. For interested readers, Figure~S3 in the Supplementary Material displays the posterior distributions of the regression coefficients. Among the main findings, the probability of being a daily smoker is lower for females, whereas higher educational levels are associated with reduced cigarette consumption.

\begin{figure}
        \centering
        \includegraphics[width=0.7\linewidth]{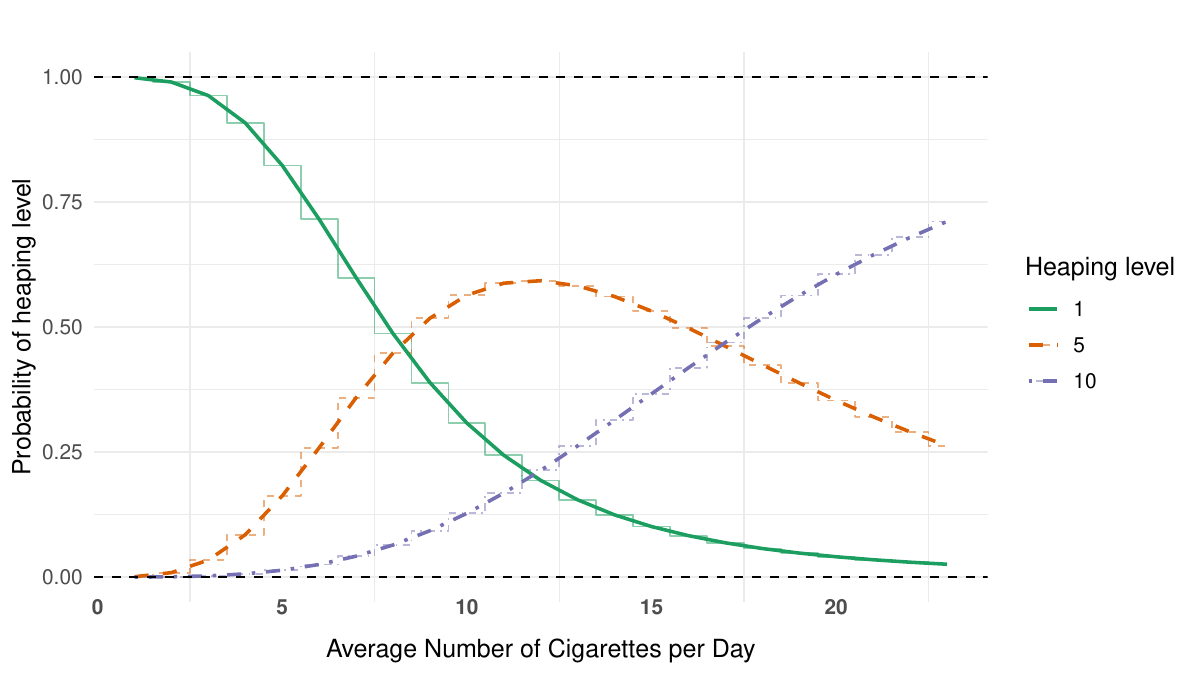}
        \caption{Relationship between the heaping level probabilities and the average number of cigarettes per day ($Z$).}
        \label{fig:app2}
    \end{figure}

Table~\ref{tab:LN_LNM} also reports posterior summaries for the parameters governing the coarsening mechanism $(\gamma_{01},\gamma_{02},\gamma_1)$, which are similar across the two specifications. In particular, $\gamma_1$ is estimated to be negative, suggesting that coarsening at multiples of 5 and 10 becomes more likely as $Z$ increases.
This pattern is illustrated in Figure~\ref{fig:app2}. At low consumption levels, responses are mostly consistent with no heaping (rounding to the nearest integer), and this probability decreases with $Z$, approaching zero around 20 cigarettes per day. Heaping at multiples of 5 is most likely at intermediate levels (roughly between 10 and 15 cigarettes per day), whereas heaping at multiples of 10 increases with $Z$ and exceeds 0.5 beyond about 15 cigarettes per day. The figure reports both the probabilities in \eqref{eq:probsG} and their discretized version in \eqref{eq:discr_prob}, which is used in model fitting.

\subsection{Small Area Estimates}\label{sec:5.2}

\begin{figure}
    \centering
    \includegraphics[width=0.7\linewidth]{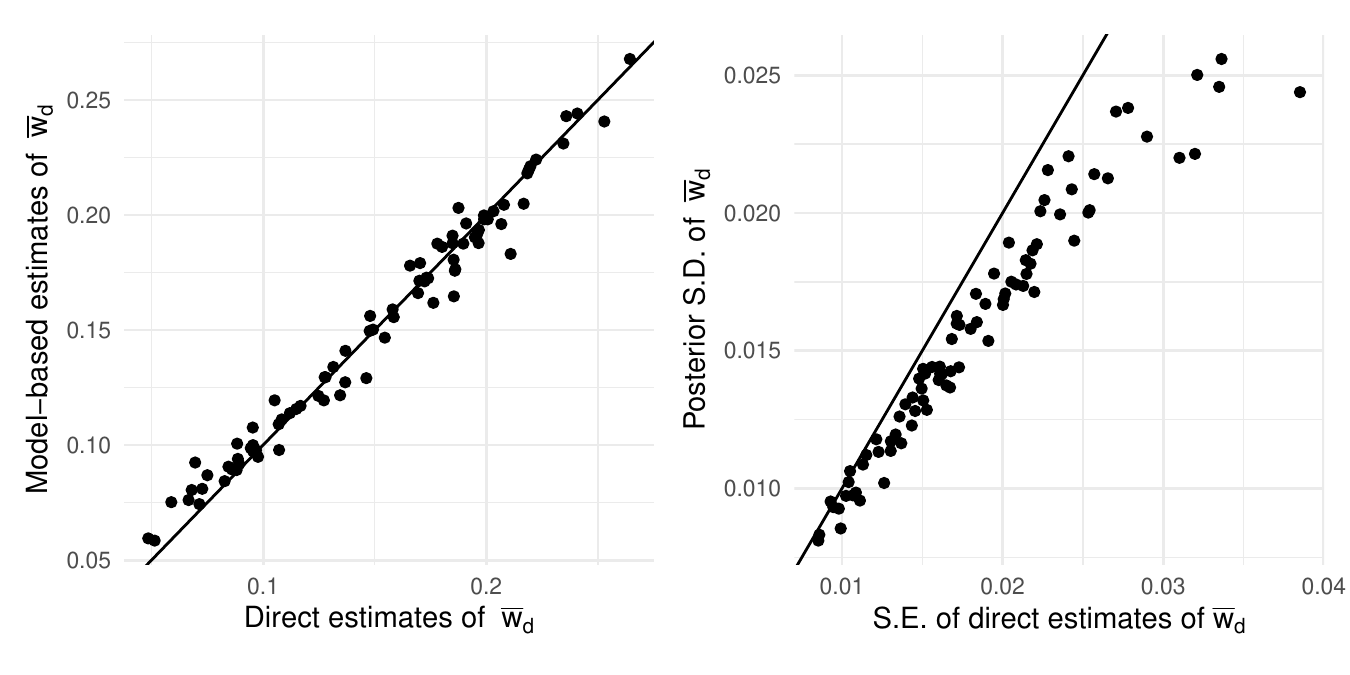}
    \caption{Comparison of design- and model-based estimates and their standard errors.}
    \label{fig:shrinkage}
\end{figure}

The analysis and interpretation of small area estimates typically begin by comparing direct and model-based estimators. However, as discussed in Section~\ref{sec:2}, in the present application the only target functional for which unbiased direct estimates are available is $\overline{w}_d$. Focusing on this indicator, Figure~\ref{fig:shrinkage} shows that model-based estimates are positively correlated with the corresponding direct estimates (left panel), while exhibiting substantially lower variability (right panel). This supports the use of the model not only for handling coarsening in the intensity component, but also for stabilizing direct estimates for region--age domains that are not explicitly planned in the survey design.

\begin{figure}
        \centering
        \includegraphics[width=\linewidth]{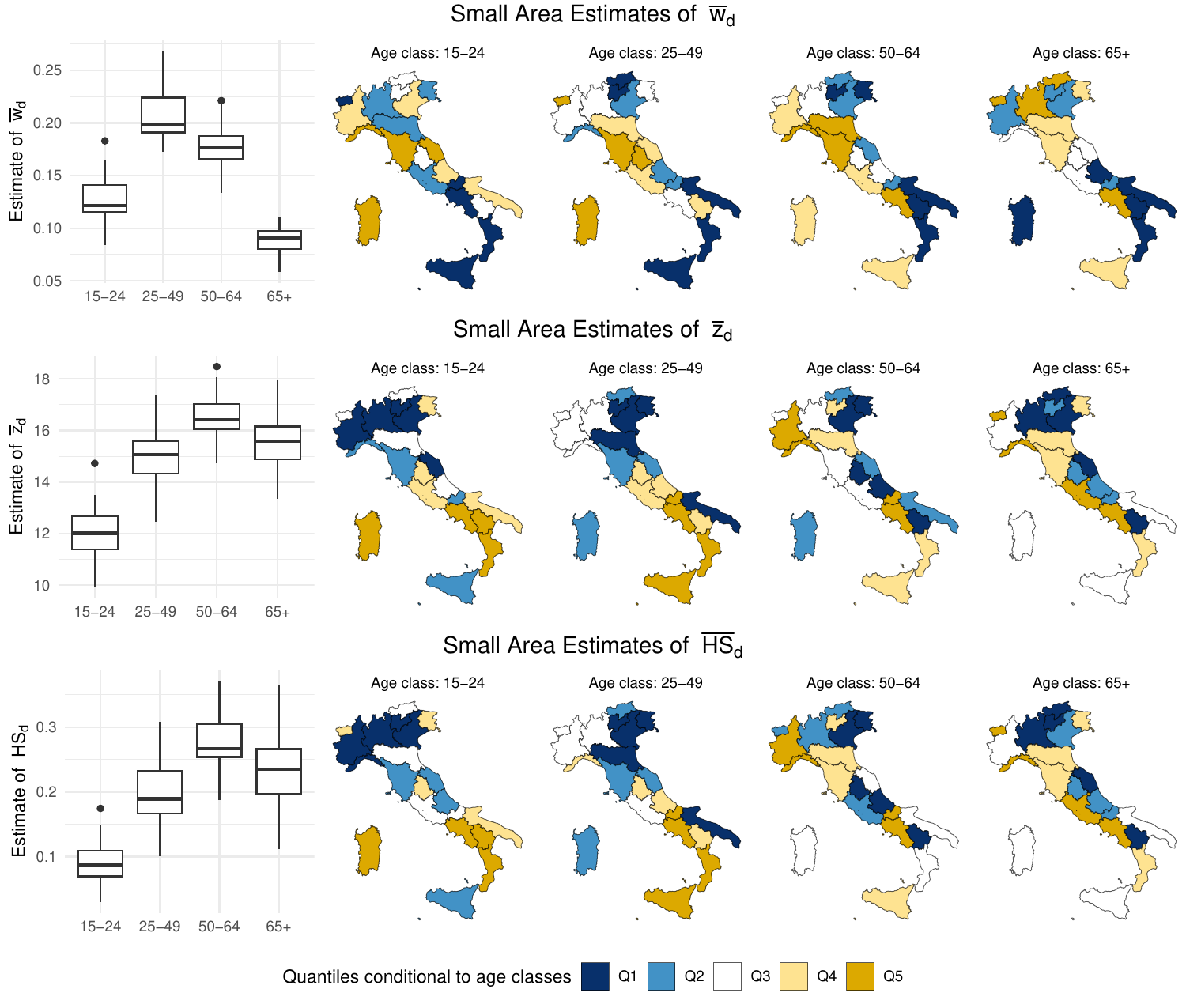}
        \caption{Box-plots of small area estimates of $\overline{w}_d$, $\bar{z}_d$, and $\overline{HS}_d$, along with the corresponding choropleth maps of age-class conditional quantiles.}
        \label{fig:maps}
    \end{figure}

We now turn to the presentation and interpretation of the estimated indicators, with the aim of extracting substantive information on the phenomenon under study for the domains of interest. To this end, Figure~\ref{fig:maps} provides insight into age-specific patterns through box plots of model-based estimates, and into the spatial distribution through choropleth maps stratified by age class, with quintiles computed within each stratum.

Focusing on the box plots in the first column of Figure~\ref{fig:maps}, we start from $\overline{w}_d$. The proportion of daily smokers is highest among individuals aged 25--49, with regional values ranging approximately between 0.173 and 0.268, and then declines at older ages. Estimates for the 15--24 group are intermediate, lying between the adult classes and the 65+ group. A different pattern emerges for $\overline{z}_d$ and $\overline{HS}_d$, reported in the second and third rows: for both indicators, the box plots display a clear age gradient, with smoking intensity and the proportion of heavy smokers increasing with age, peaking among individuals aged 50--64, and remaining at comparable levels in the 65+ class. Overall, a sizable share of the variation in the estimates reflects differences across age classes, while, within each age-group, regional values are more concentrated.

In light of these differences, the choropleth maps associated with each age class, reporting quintiles of the indicator (lower values in Q1 and higher values in Q5), help explore geographical variation across regions. For $\overline{w}_d$, the maps do not suggest clear spatial clusters or a sharply defined geographical gradient, and spatial patterns vary across age classes. Overall, lower values (Q1) are mostly observed in Southern Italy, with a few Northern regions falling in Q1 for some age classes. Conversely, higher values (Q5) are more common in Central Italy, although in the 65+ group higher values are more frequently observed in Northern Italy.
A different picture emerges from the maps of the smoking-intensity indicators $\overline{z}_d$ and $\overline{HS}_d$. Compared with $\overline{w}_d$, the spatial pattern is broadly reversed: regions in the highest quintile (Q5) are more often located in Southern Italy, whereas lower values are mainly observed in Northern regions, particularly in the North-East. Moreover, the two indicators convey consistent information, pointing to similar geographical patterns in smoking intensity.

\begin{figure}
    \centering
    \includegraphics[width=.9\linewidth]{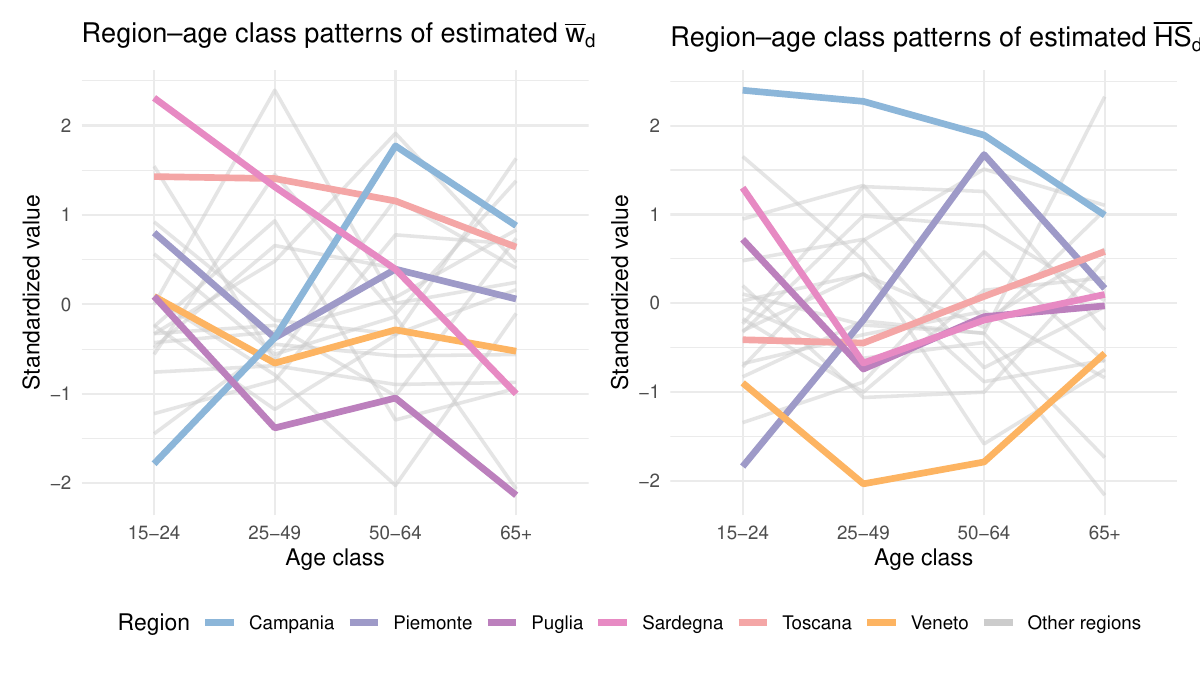}
    \caption{Parallel plots of standardized values of the estimated $\overline{w}_d$ and $\overline{HS}_d$ at the regional level across age classes.}
    \label{fig:parallel}
\end{figure}

A further perspective is to examine how each region evolves across age classes. This is done in Figure~\ref{fig:parallel}, which shows parallel plots of the standardized estimates of $\overline{w}_d$ and $\overline{HS}_d$ across age classes for each region. The indicator $\overline{z}_d$ is omitted, since it was shown above to provide information largely overlapping with $\overline{HS}_d$. Figure~\ref{fig:parallel} highlights a subset of informative regions, while trajectories for all regions are reported in Figure~S4 in the Supplementary Material.

Campania region shows relatively low standardized values of $\overline{w}_d$ in the 15--24 group, increasing with age, and this is accompanied by high standardized values of $\overline{HS}_d$, especially at younger ages. This suggests that, despite a comparatively low overall proportion of daily smokers, smoking in Campania tends to be more intense among those who do smoke. Piemonte shows a fluctuating profile of $\overline{w}_d$, remaining around the middle of the distribution, alongside a steady increase in standardized $\overline{HS}_d$, particularly from 15--24 to 50--64. Puglia is characterized by persistently low standardized values of $\overline{w}_d$ in middle-aged and older groups, mirrored by relatively low $\overline{HS}_d$ in intermediate age classes, pointing to a limited presence of heavy smokers. Sardegna exhibits a marked decline in $\overline{w}_d$ with age, while $\overline{HS}_d$ is comparatively higher among younger smokers, indicating that smoking is both more prevalent and more intense at younger ages. Toscana shows a more regular trajectory, with consistently high $\overline{w}_d$ across age classes and $\overline{HS}_d$ close to average at younger ages and increasing among older individuals. Finally, Veneto displays low standardized values of $\overline{w}_d$ across age classes, accompanied by generally low $\overline{HS}_d$, with relatively higher values at younger and older ages and a dip in middle age, suggesting that lower smoking prevalence is associated with reduced heavy smoking, particularly among middle-aged adults.


\section{Concluding remarks}\label{sec:6}

This paper proposes a Bayesian unit-level SAE framework that explicitly accounts for coarsening in self-reported health behaviours. Focusing on smoking-related indicators derived from the EHIS, we show that rounding, heaping, and top-coding can distort both point estimates and uncertainty measures when ignored. By modeling reported cigarette counts as coarsened realizations of an underlying latent continuous variable within a two-part SAE model, and by adopting a mixture-based specification for the positive component, the proposed approach delivers stable and interpretable estimates of smoking indicators. Although the model is non-trivial, to facilitate its use in practice we release the \texttt{Stan} code implementing the proposed specification, which can be adapted by practitioners.

The simulation study indicates that ignoring coarsening may lead to bias, inflated mean squared error, and undercoverage, particularly for tail-sensitive indicators such as the proportion of heavy smokers. In contrast, the coarsening-aware mixture specification improves accuracy and yields coverage closer to the nominal level across a range of realistic scenarios. The empirical application to Italian region--age domains further illustrates the practical relevance of the approach, showing how accounting for coarsening and latent heterogeneity can affect inference and support more targeted public health assessments, with model adequacy further supported by posterior predictive checks reported in Section~\ref{sec:5.1}.

Several extensions warrant further investigation. In the proposed specification, the positive component is modeled through a two-component Lognormal mixture. While, in our application, additional mixture components did not materially affect the target domain estimates, in other settings it may be useful to develop principled strategies for selecting the number of components. More broadly, although the coarsening model is tailored to the EHIS smoking questions and therefore relies on assumptions that are specific to the available response format, the underlying approach is readily extendable to other outcomes affected by heaping and/or top-coding by adapting the latent coarsening mechanism to the relevant questionnaire design. Remaining in the survey and small area setting, future research could also explore design-consistent or calibration-based approaches to obtain unbiased estimators for coarsened and top-coded variables, which could then be integrated into SAE procedures that accommodate auxiliary covariates in a more direct area-level formulation.

\end{document}